\documentclass[aps,twocolumn,showpacs,amsmath,amssymb,prb,reprint]{revtex4-1} 
\usepackage{graphicx}%
\usepackage{dcolumn}
\usepackage{bm}
\usepackage{braket}
\usepackage{color}
\begin{document} 


\title{Wavelength-dependence of laser excitation process on silicon surface}
\author{T.Otobe}
\affiliation{National Institutes for Quantum and Radiological Science and Technology, Kyoto 619-0215, Japan}
\begin{abstract}
We report a first-principle calculation for the wavelength-dependence of a laser excitation process on a silicon surface.
Although  lower frequency laser is reflected by a lower density plasma, 
it can penetrate thicker plasma sheet. 
Therefore, the depth of the laser processing depends on the width of the plasma at the surface and laser wavelength. 
The time-dependent density-functional theory and Maxwell’s equations are simultaneously employed to elucidate the effect of laser propagation on laser-matter interaction under ultrafast pulse lasers (FWHM:12 fs).
A longer-wavelength laser field facilitates deeper melting and ablation in silicon, despite a lower critical plasma density. 
Such a deeper excitation by a longer wavelength is because of the penetration of the laser field through the plasma on the surface. 
The plasma-formation depth is saturated at approximately half the wavelength in silicon.
\end{abstract}
\maketitle
\section{introduction}

Processing of solid-state materials using femtosecond laser pulses
has attracted considerable interest because of their potential applications
in high-precision processing technology. 
\cite{Chichkov,Stuart,Liu,Lenzner,Geissler,Lenzner00,Sudrie,Doumy,Amoruso,Gattass08, Gamaly11,Chimier}

In particular, a pulse with a duration of few tens of femtoseconds (fs) enables the processing of dielectric surfaces without thermal damage because this duration is considerably shorter than the thermalization duration (ps$\sim$)\cite{Sundaram02, Gattass08}.
Moreover, the excitation of dielectrics by intense laser fields is employed in plasma optics such as plasma mirrors \cite{Doumy, Tsubouchi12}.

Many experimental and theoretical works on laser processing using near infrared (NIR) lasers have been reported.
However, recently, the progress of laser technology has made intense mid-IR (MIR) lasers available \cite{Ghimire11,Austin15}.
As the photon energy of the IR laser is considerably lower than the dielectric bandgap, nonlinear excitation (multiphoton absorption and tunnel ionization) is a critical process.
In general, a theoretical treatment for laser-electron nonlinear interaction is described by the rate equation that includes the electron excitation by the Keldysh theory\cite{Keldysh}, the avalanche effect, Joule heating, and the Drude model \cite{Stuart}.


The time-evolution of the laser field can be described by Maxwell’s equations, considering the material properties through the constitutive relations. 
For ordinary light pulses, the response of the medium is linear in the electromagnetic field, and it is characterized by linear susceptibilities.
However, for intense and ultrashort laser pulses, conditions that require theoretical treatment beyond the linear response are encountered. 
If perturbative expansion is no longer useful, the time-dependent Schrodinger equation must be applied for electrons and solved in the time domain.

For a many-electron system, the time-dependent density-functional theory (TDDFT) \cite{Runge84} has been applied for laser-molecule and solid state interaction \cite{Tong01, Otobe08}.
We consider the TDDFT as the only \textit{ab initio} quantum method applicable to strong electromagnetic fields in condensed media. 
In our previous work, we developed a formalism and computational method to describe the propagation of an intense electromagnetic field in a condensed medium, incorporating the electron dynamics feedback to the electromagnetic field \cite{yabana12, Sato15,salmon}.
In particular, in the case of processing with an MIR laser,  the plasma frequency of the excited electron-hole pairs affects the dynamics of the 
electromagnetic field at lower plasma density because the plasma density easily reaches the frequency of an MIR laser. 
Although a lower frequency laser is reflected by the plasma at the surface with a lower laser intensity,
a wider plasma sheet must be formed to reflect the MIR laser.
Therefore, we must clarify the relationship between the excitation depth and laser frequency to understand laser processing under various laser frequencies.

In this study, we present the first-principle simulation of the laser frequency and intensity-dependence of the laser excitation process on silicon surface employing the above multiscale approach.
We assume an ultrafast pulse laser (12 fs FWHM) in the MIR--NIR frequencies.
We present the position dependence of the excitation energy and electron-hole density with various laser parameters to elucidate the laser excitation process and plasma-mirror formation at the surface.

 The remainder of this paper is organized as follows:
 In section II, we describe our first-principle multiscale formalism to calculate the laser-matter interaction on the surface.
 In section III, we present the numerical results.
 In section IV, we summarize the study. 
 
\section{Computational Method}
As the theory and its implementation employed in the 
calculation herein are described elsewhere \cite{Bertsch00, yabana12,Sato15}, we explain it in brief.
The laser pulse that enters from a vacuum and
attenuates in the medium varies on a micrometer scale,
whereas the electron dynamics occur on a subnanometer
scale. To overcome these conflicting spatial scales, we 
develop a multiscale implementation, introducing two coordinate
systems: a macroscopic coordinate $X$ for laser pulse
propagation and a microscopic coordinate $r$ for the local electron
dynamics. The laser pulse is described by the vector potential $\vec{A}_X (t)$ which satisfies
\begin{equation}
\frac{1}{c^2}\frac{\partial^2 \vec{A}_X (t)}{\partial t^2}-\frac{\partial^2 \vec{A}_X (t)}{\partial X^2} =-\frac{4\pi e^2}{c}\vec{J}_X\left(t\right).
\label{MXE}
\end{equation}

At each point $X$, we consider the lattice-periodic electron dynamics
driven by electric field $E_X(t)=-\frac{1}{c}dA_X(t)/dt$. They
are described by the electron orbitals $\psi_{i,X}(\vec{r},t)$ which satisfy
the time-dependent Kohn-Sham equation
\begin{eqnarray}
i\hbar \frac{\partial}{\partial t} \psi_{i,X}\left(\vec{r},t\right)&=&\Bigg[ \frac{1}{2m} \left( -i\hbar\nabla_r+\frac{e}{c}\vec{A}_X\left(t\right)\right)^2-\phi_X\left(\vec{r},t\right)\nonumber\\
&+&\mu_{xc,X}\left(\vec{r},t\right)\Bigg] \psi_{i,X}\left(\vec{r},t\right),
\label{TDKS}
\end{eqnarray}
where the potential $\phi_X(\vec{r},t)$ which includes the Hartree and
ionic contributions, and the exchange-correlation potential $\mu_{xc,X}(\vec{r},t)$, are periodic in the lattice. The electric current $J_X(t)$
is provided from the electron orbitals:
\begin{eqnarray} 
J_X\left(t\right) &=& -\frac{e}{mV} \int_{V} d\vec r \sum_i 
{\rm Re} \psi_{i,X}^* \left( \vec p + \frac{e}{c}\vec A_X(t) \right) \psi_{i,X} \nonumber\\
&+& J_{X,NL}(t), 
 \label{CUR} 
\end{eqnarray} 
where $V$ is the volume of a unit cell;
$J_{X,NL}(t)$ is the current caused by the nonlocality of the pseudopotential. 

We solve Eqs.~(\ref{MXE})--(\ref{CUR})
simultaneously as an initial value problem, where the incident
laser pulse is prepared in a vacuum region on the top of the
surface, while all the Kohn-Sham orbitals are set to their ground
states.
In this study, we use the modified
Becke-Johnson exchange potential (mBJ) \cite{Becke2006} specified in
Ref.~\cite{mBJ2} (Eqs. (2)--(4)) with an LDA correlation potential \cite{LDA} in the
adiabatic approximation.
\begin{figure}
\includegraphics[width=90mm]{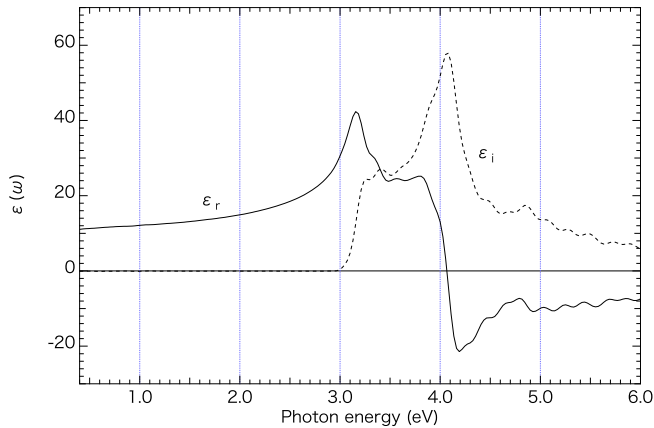}
\caption{\label{fig:Fig0} Real and imaginary parts of the dielectric function calculated by TDDFT with the mBJ potential. } 
\end{figure}
The mBJ potential improves the bandgap which is under-estimated in conventional LDA.
Figure \ref{fig:Fig0} depicts the real and imaginary parts of the dielectric function calculated by TDDFT with the mBJ as a function of the photon energy. The calculated optical bandgap is 3.0 eV, which is an improvement compared to that obtained using the LDA (2.4 eV) \cite{yabana12}, and it is approximately equal to the experimental value (3.1 eV). 

Our multiscale calculation uses a one-dimensional grid with a spacing of 250 atomic units for the propagation of laser electromagnetic fields. 
At each grid point, the electron dynamics are calculated using an atomic-scale cubic unit cell containing eight silicon atoms which are discretized into $24^3$ Cartesian grids. 
We discretize the Bloch momentum space into $8^3$ $k$ points.
The dynamics of the 32 valence electrons are
treated explicitly; the effects of the core electrons are considered through the pseudopotentials \cite{TM91,Kleinman82}. The electromagnetic
fields as well as electrons are evolved with a common time step of 0.04 atomic units.
Note that we discretize the Bloch momentum space into $16^3$ $k$ points for the calculation of $\varepsilon (\omega )$ (Fig.~\ref{fig:Fig0}) to obtain a smooth spectrum.
For multiscale calculation, we use a sparse $k$ grid because of the limitation of the computational resource.
Although this calculation may not provide fully convergent results, we do not expect the truncation to affect the physical results by more than 10\% \cite{yabana12}.

The incident laser field $E_{in}(X,t)$ in vacuum is 
\begin{equation}
E_{in}(X,t)=
\begin{cases}
E_0 \sin^2\left(\pi \frac{t_X}{T_p}\right)\cos(\omega_0 t_X) & 0<t_X<T_p \\
0& T_p<t_X< T_e, \label{eq:field}
\end{cases}
\end{equation} 
where $E_0$ is the peak electric-field amplitude, $\omega_0$ is the laser frequency, and $t_X=t-X/c$ describes the space-time dependence of the field.
The pulse length $T_p$ is set to 31.2 fs, and the computation is terminated at $T_e=48.3$ fs.

\section{Results and discussion}
\begin{figure}
\includegraphics[width=85mm]{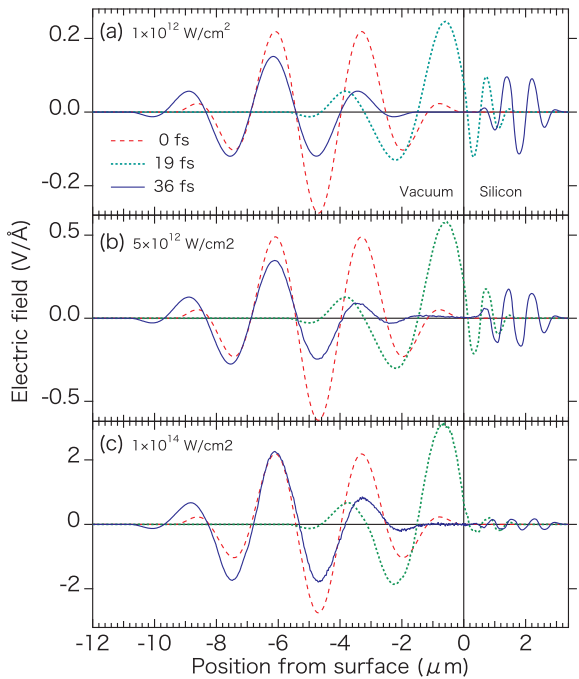}
\caption{\label{fig:Fig1} Time-evolution of the electromagnetic field for a laser frequency of 0.4 eV at intensities of (a) $1\times10^{12}$, (b) $5\times10^{12}$, and (c) $1\times10^{14}$ W/cm$^2$. } 
\end{figure}

Figure \ref{fig:Fig1} shows the time-evolution of the electromagnetic field of the pulse laser whose frequency is 0.4 eV, around the silicon surface. 
The initial field (0 fs) is denoted by a red-dashed line.
The laser collides with the silicon surface exhibiting reflection and transmission; these are denoted by a green-dotted line(19 fs).
The blue-solid line denotes the field after laser-silicon interaction.
The laser frequency is set to 0.4 eV, and the laser intensities are set to $1\times10^{12}$ (Fig.~\ref{fig:Fig1} (a)), $5\times10^{12}$ (Fig.~\ref{fig:Fig1} (b)), and $1\times10^{14}$ W/cm$^2$ (Fig.~\ref{fig:Fig1} (c)), respectively.
For the least intensity (Fig.~\ref{fig:Fig1} (a)), the reflection and transmission occur as linear processes.
Therefore, the laser field in the silicon ($X>0$) at 36 fs (blue-solid line) shows a profile similar to that of the incident field (red-dashed line for $X < 0$).

The reflectivity is defined by
\begin{equation}
R=\frac{\int_{-\infty}^0 dX |E(X,t=T_e)|^2}{\int_{-\infty}^0 dX |E_{in}(X,t=0)|^2},
\end{equation}
where $E$ is the electric field associated with the pulse.
$R$ is approximately 0.299, which is consistent with the reflectivity (0.30) calculated using the dielectric function ($\varepsilon(\omega=0.4)=11.5$). 
Meanwhile, as the laser intensity increases, the profile of the laser field in silicon is deformed to be rectangular and the reflectivity increases because photoabsorption occurs dominantly around the pulse peak. 
Reflection by the electron-hole plasma at the surface occurs, and it is considerable at the highest intensity.

The laser intensity and frequency-dependence of the reflectivity are depicted in Fig.~\ref{Fig3}.
We assume three different frequencies: 0.4 eV, 0.775 eV, and 1.55 eV.
In general, the dielectric function $\varepsilon_{Si}(\omega)$ is modulated by the plasma response.
The reflectivity is minimized when the screened plasma frequency at the surface coincides with the laser frequency.
A dip in the reflectivity can be observed at intensities of $3\times 10^{12}$, $5\times 10^{12}$, and $7\times 10^{12}$ W/cm$^2$, for $\omega_0= 0.4$ eV, 0.775 eV, and 1.55 eV respectively.
Above these critical intensities, the reflectivity increases up to 0.77 at the maximum intensity because of the metallic response of the plasma.

The laser-intensity dependencies of the reflectivity for each frequency exhibit qualitative differences. 
The increase in reflectivity becomes more moderate for lower frequencies, above the critical intensity.
This frequency dependence on the reflectivity indicates that a lower- frequency laser field can penetrate the plasma formed on the silicon surface. 
\begin{figure}
\includegraphics[width=85mm]{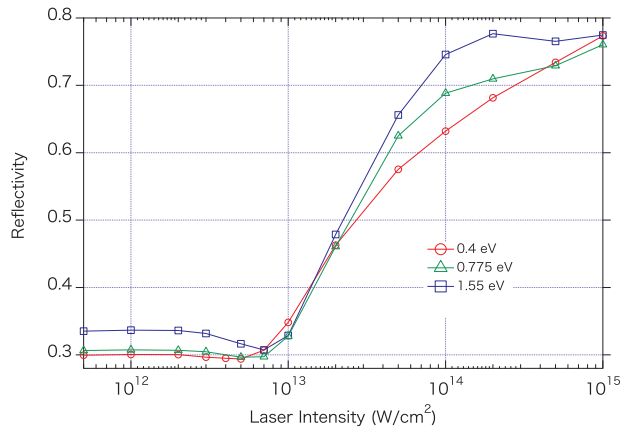}
\caption{\label{Fig2} Incident laser reflectivity at frequencies of 0.4 eV (red circle), 0.775 eV (green triangle), and 1.55 eV (blue square), respectively, as a function of the incident laser intensity.}
\end{figure}

The deposited energy per silicon atom is shown in Fig.~\ref{Fig2} as a function of the depth from the surface.
We compute the energy transfer to the medium from the
electromagnetic-side because calculation using the Kohn-
Sham densities requires an explicit energy density
functional for the mBJ potential. The energy transfer rate $W$ is given by
$W=-\vec{E}\cdot \vec{J}$,
where $E$ is the electric field associated with the pulse. 
The deposited energy density is given by 
$E_{ab}(X)=-\int dt W_{X}(t)$. 

The two dashed-black lines denote the melting and cohesive energy of silicon.
For laser processing, the melting energy is an important index, whereas, the cohesive energy indicates the ablation threshold \cite{Sato15}.
Our results clearly indicate that lower-frequency laser can excite silicon more deeply;
this is not surprising, if we consider the propagation of the long wavelength, 
although a shallower excitation depth is expected for lower frequency because a higher multiphoton process is needed for the electron excitation process at lower frequency. 
The longer-wavelength light field can penetrate the thin plasma at the surface, when the thickness of the plasma sheet is considerably smaller than the wavelength in silicon.
 
\begin{figure}
\includegraphics[width=85mm]{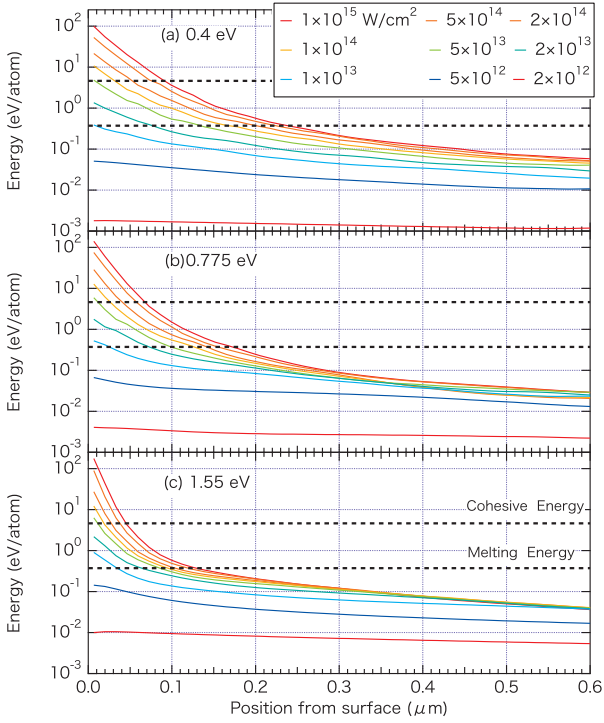}
\caption{\label{Fig3} Laser-intensity dependence of the deposited energy $E_{ab}(X)$ at (a) 0.4 eV, (b) 0.775 eV, and (c) 1.55 eV.}
\end{figure}

\begin{figure}
\includegraphics[width=85mm]{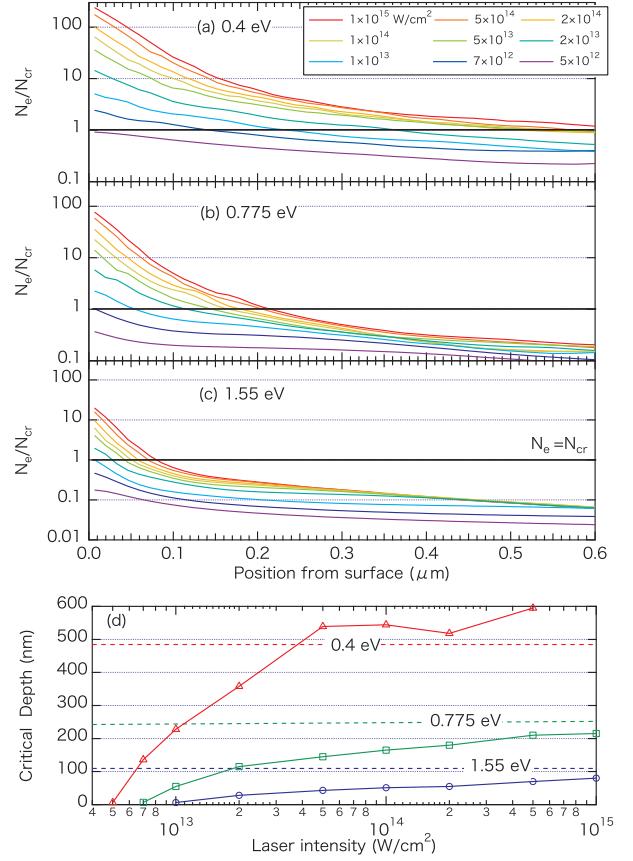}
\caption{\label{fig:Fig4} Laser-intensity dependence of the number of excited electrons ($N_e(X)$) scaled by the critical electron number ($N_{cr}$) at laser frequencies of (a) 0.4 eV, (b) 0.775 eV, and (c) 1.55 eV.
(d) Critical depth $N_e(X_{cr})=N_{cr}$ for each frequency. The dashed lines indicate half of the wavelength for each frequency in silicon.}
\end{figure}

From Fig.~\ref{Fig3}, the plasma functions as a mirror, above the critical intensities.
However, the quality of the plasma mirror depends on the laser frequency.
To elucidate the functioning of the plasma sheet as a mirror, and laser penetration into the plasma sheet, we depict the position-dependent electron-hole density ($N_e$) in Fig.~\ref{fig:Fig4}.
The laser-intensity dependencies of $N_e$ as a function of the position from the surface are shown in Figs.~\ref{fig:Fig4}(a)--(c).
$N_e$ is defined as the projection of $\psi_{i,X}$ to the ground states at the same vector potential ($\Phi_{i,X}$):
\begin{equation}
N_{e}(X) = \frac{1}{V}\sum_{ii' =occ} \left( \delta_{ii'} -  
\vert \langle \Phi_{i,X} \vert \psi_{i',X}(t=T_e) \rangle \vert^2 \right),
\label{eq:nex}
\end{equation}
at the end of the time-evolution.
We scale $N_e$ using the critical density for the screened plasma:
\begin{equation}
N_{cr}\equiv \frac{\omega_0^2 m^* \varepsilon(\omega_0)}{4\pi e^2},
\end{equation}
where $m^*=0.3m$ is the effective mass, $\varepsilon(\omega_0)$ is the dielectric function at frequency of $\omega_0$.
$N_{cr}$ is 0.008/atom for 0.4 eV, 0.033/atom for 0.775 eV, and 0.15/atom for 1.55 eV.
Fig.~\ref{fig:Fig4} (d) shows the critical depth ($X_{cr}$) defined by the position $N_e(X_{cr})=N_{cr}$ as a function of the laser intensity.
Around the critical intensity, $X_{cr}$ corresponds to the surface ($X\sim0$) at all the frequencies.
It should be noted that $N_e$ should be defined in the absence of an electric field because it modulates the ground state.
In our calculation, the very-weak light field induced by the spontaneous oscillation of $J_X(t)$ renders $N_e$ larger than the ideal value.
In particular, in the case of 0.4 eV, whose $N_{cr}$ is very small, $X_{cr}$ is considered as the reference.
However, our results provide sufficient accuracy for $X_{cr}$ to discuss the plasma thickness \cite{yabana12}. 

With the increase in laser intensity, $X_{cr}$ increases rapidly, immediately above the critical intensities (Fig.~\ref{fig:Fig4} (d)).
However, $X_{cr}$ is saturated at approximately 500 nm for 0.4 eV, 200 nm for 0.775 eV, and 100 nm for 1.55 eV.
The saturated positions ($X_{cr}^{sat}$) correspond to half the wavelength in silicon (dashed lines) for all the frequencies. 
These results demonstrate that the functioning of the plasma mirror is inadequate until the plasma thickness increases sufficiently.
The saturation of the reflectivity shown in Fig.~\ref{Fig2} and $X_{cr}$ (Fig.~\ref{fig:Fig4}) indicate that 
the plasma reflection is saturated, when $X_{cr}$ becomes quarter of the wavelength in silicon, in the case of 1.55 and 0.775 eV. 
After the plasma reflection is saturated, the thickness of the plasma accesses the $X_{cr}^{sat}$ value.

\section{Conclusion}
In summary, a first-principle simulation of the laser-intensity dependence on the processing of the silicon-surface was presented in this study.
The obtained results indicate that low-frequency (long-wavelength) laser can excite deeper layers.
The frequency dependence of the laser-processing depth is attributed to the ratio between the thickness of the plasma sheet formed at the surface.
The functioning of the formed plasma as a mirror is inadequate, 
until the thickness of the plasma becomes comparable to quarter of the wavelength in silicon.
In particular, the thickness of the plasma accesses half the wavelength in silicon with an increase in laser intensity. 
It may be possible to optimize the characteristic depth of ablation and/or the melting of silicon using lasers of various frequencies.
The wavelength-dependence of the quality of the plasma mirror provides crucial insights for plasma optics. 

\section*{Acknowledgement}
This work is supported by MEXT Q-LEAP and JSPS KAKENHI (Grant No.~JP17H03525). 
The numerical calculations were performed on supercomputer SGI ICE X at 
the Japan Atomic Energy Agency (JAEA).

\bibliography{LaserProc2.bib}

\begin{thebibliography}{29}%
\makeatletter
\providecommand \@ifxundefined [1]{%
 \@ifx{#1\undefined}
}%
\providecommand \@ifnum [1]{%
 \ifnum #1\expandafter \@firstoftwo
 \else \expandafter \@secondoftwo
 \fi
}%
\providecommand \@ifx [1]{%
 \ifx #1\expandafter \@firstoftwo
 \else \expandafter \@secondoftwo
 \fi
}%
\providecommand \natexlab [1]{#1}%
\providecommand \enquote  [1]{``#1''}%
\providecommand \bibnamefont  [1]{#1}%
\providecommand \bibfnamefont [1]{#1}%
\providecommand \citenamefont [1]{#1}%
\providecommand \href@noop [0]{\@secondoftwo}%
\providecommand \href [0]{\begingroup \@sanitize@url \@href}%
\providecommand \@href[1]{\@@startlink{#1}\@@href}%
\providecommand \@@href[1]{\endgroup#1\@@endlink}%
\providecommand \@sanitize@url [0]{\catcode `\\12\catcode `\$12\catcode
  `\&12\catcode `\#12\catcode `\^12\catcode `\_12\catcode `\%12\relax}%
\providecommand \@@startlink[1]{}%
\providecommand \@@endlink[0]{}%
\providecommand \url  [0]{\begingroup\@sanitize@url \@url }%
\providecommand \@url [1]{\endgroup\@href {#1}{\urlprefix }}%
\providecommand \urlprefix  [0]{URL }%
\providecommand \Eprint [0]{\href }%
\providecommand \doibase [0]{https://doi.org/}%
\providecommand \selectlanguage [0]{\@gobble}%
\providecommand \bibinfo  [0]{\@secondoftwo}%
\providecommand \bibfield  [0]{\@secondoftwo}%
\providecommand \translation [1]{[#1]}%
\providecommand \BibitemOpen [0]{}%
\providecommand \bibitemStop [0]{}%
\providecommand \bibitemNoStop [0]{.\EOS\space}%
\providecommand \EOS [0]{\spacefactor3000\relax}%
\providecommand \BibitemShut  [1]{\csname bibitem#1\endcsname}%
\let\auto@bib@innerbib\@empty
\bibitem [{\citenamefont {Chichkov}\ \emph {et~al.}(1996)\citenamefont
  {Chichkov}, \citenamefont {Momma}, \citenamefont {Nolte}, \citenamefont {von
  Alvensleben},\ and\ \citenamefont {T{\"u}nnermann}}]{Chichkov}%
  \BibitemOpen
  \bibfield  {author} {\bibinfo {author} {\bibfnamefont {B.~N.}\ \bibnamefont
  {Chichkov}}, \bibinfo {author} {\bibfnamefont {C.}~\bibnamefont {Momma}},
  \bibinfo {author} {\bibfnamefont {S.}~\bibnamefont {Nolte}}, \bibinfo
  {author} {\bibfnamefont {F.}~\bibnamefont {von Alvensleben}},\ and\ \bibinfo
  {author} {\bibfnamefont {A.}~\bibnamefont {T{\"u}nnermann}},\ }\bibfield
  {title} {\bibinfo {title} {Femtosecond, picosecond and nanosecond laser
  ablation of solids},\ }\href {https://doi.org/10.1007/BF01567637} {\bibfield
  {journal} {\bibinfo  {journal} {Applied Physics A}\ }\textbf {\bibinfo
  {volume} {63}},\ \bibinfo {pages} {109} (\bibinfo {year} {1996})}\BibitemShut
  {NoStop}%
\bibitem [{\citenamefont {Stuart}\ \emph {et~al.}(1996)\citenamefont {Stuart},
  \citenamefont {Feit}, \citenamefont {Herman}, \citenamefont {Rubenchik},
  \citenamefont {Shore},\ and\ \citenamefont {Perry}}]{Stuart}%
  \BibitemOpen
  \bibfield  {author} {\bibinfo {author} {\bibfnamefont {B.~C.}\ \bibnamefont
  {Stuart}}, \bibinfo {author} {\bibfnamefont {M.~D.}\ \bibnamefont {Feit}},
  \bibinfo {author} {\bibfnamefont {S.}~\bibnamefont {Herman}}, \bibinfo
  {author} {\bibfnamefont {A.~M.}\ \bibnamefont {Rubenchik}}, \bibinfo {author}
  {\bibfnamefont {B.~W.}\ \bibnamefont {Shore}},\ and\ \bibinfo {author}
  {\bibfnamefont {M.~D.}\ \bibnamefont {Perry}},\ }\bibfield  {title} {\bibinfo
  {title} {Nanosecond-to-femtosecond laser-induced breakdown in dielectrics},\
  }\href {https://doi.org/10.1103/PhysRevB.53.1749} {\bibfield  {journal}
  {\bibinfo  {journal} {Phys. Rev. B}\ }\textbf {\bibinfo {volume} {53}},\
  \bibinfo {pages} {1749} (\bibinfo {year} {1996})}\BibitemShut {NoStop}%
\bibitem [{\citenamefont {Liu}\ \emph {et~al.}(1997)\citenamefont {Liu},
  \citenamefont {Du},\ and\ \citenamefont {Mourou}}]{Liu}%
  \BibitemOpen
  \bibfield  {author} {\bibinfo {author} {\bibfnamefont {X.}~\bibnamefont
  {Liu}}, \bibinfo {author} {\bibfnamefont {D.}~\bibnamefont {Du}},\ and\
  \bibinfo {author} {\bibfnamefont {G.}~\bibnamefont {Mourou}},\ }\bibfield
  {title} {\bibinfo {title} {Laser ablation and micromachining with ultrashort
  laser pulses},\ }\href {https://doi.org/10.1109/3.631270} {\bibfield
  {journal} {\bibinfo  {journal} {IEEE Journal of Quantum Electronics}\
  }\textbf {\bibinfo {volume} {33}},\ \bibinfo {pages} {1706} (\bibinfo {year}
  {1997})}\BibitemShut {NoStop}%
\bibitem [{\citenamefont {Lenzner}\ \emph {et~al.}(1998)\citenamefont
  {Lenzner}, \citenamefont {Kr\"uger}, \citenamefont {Sartania}, \citenamefont
  {Cheng}, \citenamefont {Spielmann}, \citenamefont {Mourou}, \citenamefont
  {Kautek},\ and\ \citenamefont {Krausz}}]{Lenzner}%
  \BibitemOpen
  \bibfield  {author} {\bibinfo {author} {\bibfnamefont {M.}~\bibnamefont
  {Lenzner}}, \bibinfo {author} {\bibfnamefont {J.}~\bibnamefont {Kr\"uger}},
  \bibinfo {author} {\bibfnamefont {S.}~\bibnamefont {Sartania}}, \bibinfo
  {author} {\bibfnamefont {Z.}~\bibnamefont {Cheng}}, \bibinfo {author}
  {\bibfnamefont {C.}~\bibnamefont {Spielmann}}, \bibinfo {author}
  {\bibfnamefont {G.}~\bibnamefont {Mourou}}, \bibinfo {author} {\bibfnamefont
  {W.}~\bibnamefont {Kautek}},\ and\ \bibinfo {author} {\bibfnamefont
  {F.}~\bibnamefont {Krausz}},\ }\bibfield  {title} {\bibinfo {title}
  {Femtosecond optical breakdown in dielectrics},\ }\href
  {https://doi.org/10.1103/PhysRevLett.80.4076} {\bibfield  {journal} {\bibinfo
   {journal} {Phys. Rev. Lett.}\ }\textbf {\bibinfo {volume} {80}},\ \bibinfo
  {pages} {4076} (\bibinfo {year} {1998})}\BibitemShut {NoStop}%
\bibitem [{\citenamefont {Geissler}\ \emph {et~al.}(1999)\citenamefont
  {Geissler}, \citenamefont {Tempea}, \citenamefont {Scrinzi}, \citenamefont
  {Schn\"urer}, \citenamefont {Krausz},\ and\ \citenamefont
  {Brabec}}]{Geissler}%
  \BibitemOpen
  \bibfield  {author} {\bibinfo {author} {\bibfnamefont {M.}~\bibnamefont
  {Geissler}}, \bibinfo {author} {\bibfnamefont {G.}~\bibnamefont {Tempea}},
  \bibinfo {author} {\bibfnamefont {A.}~\bibnamefont {Scrinzi}}, \bibinfo
  {author} {\bibfnamefont {M.}~\bibnamefont {Schn\"urer}}, \bibinfo {author}
  {\bibfnamefont {F.}~\bibnamefont {Krausz}},\ and\ \bibinfo {author}
  {\bibfnamefont {T.}~\bibnamefont {Brabec}},\ }\bibfield  {title} {\bibinfo
  {title} {Light propagation in field-ionizing media: Extreme nonlinear
  optics},\ }\href {https://doi.org/10.1103/PhysRevLett.83.2930} {\bibfield
  {journal} {\bibinfo  {journal} {Phys. Rev. Lett.}\ }\textbf {\bibinfo
  {volume} {83}},\ \bibinfo {pages} {2930} (\bibinfo {year}
  {1999})}\BibitemShut {NoStop}%
\bibitem [{\citenamefont {Lenzner}\ \emph {et~al.}(2000)\citenamefont
  {Lenzner}, \citenamefont {Krausz}, \citenamefont {Kr{\"u}ger},\ and\
  \citenamefont {Kautek}}]{Lenzner00}%
  \BibitemOpen
  \bibfield  {author} {\bibinfo {author} {\bibfnamefont {M.}~\bibnamefont
  {Lenzner}}, \bibinfo {author} {\bibfnamefont {F.}~\bibnamefont {Krausz}},
  \bibinfo {author} {\bibfnamefont {J.}~\bibnamefont {Kr{\"u}ger}},\ and\
  \bibinfo {author} {\bibfnamefont {W.}~\bibnamefont {Kautek}},\ }\bibfield
  {title} {\bibinfo {title} {Photoablation with sub-10 fs laser pulses},\
  }\href {https://doi.org/https://doi.org/10.1016/S0169-4332(99)00432-8}
  {\bibfield  {journal} {\bibinfo  {journal} {Applied Surface Science}\
  }\textbf {\bibinfo {volume} {154-155}},\ \bibinfo {pages} {11 } (\bibinfo
  {year} {2000})}\BibitemShut {NoStop}%
\bibitem [{\citenamefont {Sudrie}\ \emph {et~al.}(2002)\citenamefont {Sudrie},
  \citenamefont {Couairon}, \citenamefont {Franco}, \citenamefont {Lamouroux},
  \citenamefont {Prade}, \citenamefont {Tzortzakis},\ and\ \citenamefont
  {Mysyrowicz}}]{Sudrie}%
  \BibitemOpen
  \bibfield  {author} {\bibinfo {author} {\bibfnamefont {L.}~\bibnamefont
  {Sudrie}}, \bibinfo {author} {\bibfnamefont {A.}~\bibnamefont {Couairon}},
  \bibinfo {author} {\bibfnamefont {M.}~\bibnamefont {Franco}}, \bibinfo
  {author} {\bibfnamefont {B.}~\bibnamefont {Lamouroux}}, \bibinfo {author}
  {\bibfnamefont {B.}~\bibnamefont {Prade}}, \bibinfo {author} {\bibfnamefont
  {S.}~\bibnamefont {Tzortzakis}},\ and\ \bibinfo {author} {\bibfnamefont
  {A.}~\bibnamefont {Mysyrowicz}},\ }\bibfield  {title} {\bibinfo {title}
  {Femtosecond laser-induced damage and filamentary propagation in fused
  silica},\ }\href {https://doi.org/10.1103/PhysRevLett.89.186601} {\bibfield
  {journal} {\bibinfo  {journal} {Phys. Rev. Lett.}\ }\textbf {\bibinfo
  {volume} {89}},\ \bibinfo {pages} {186601} (\bibinfo {year}
  {2002})}\BibitemShut {NoStop}%
\bibitem [{\citenamefont {Doumy}\ \emph {et~al.}(2004)\citenamefont {Doumy},
  \citenamefont {Qu\'er\'e}, \citenamefont {Gobert}, \citenamefont {Perdrix},
  \citenamefont {Martin}, \citenamefont {Audebert}, \citenamefont {Gauthier},
  \citenamefont {Geindre},\ and\ \citenamefont {Wittmann}}]{Doumy}%
  \BibitemOpen
  \bibfield  {author} {\bibinfo {author} {\bibfnamefont {G.}~\bibnamefont
  {Doumy}}, \bibinfo {author} {\bibfnamefont {F.}~\bibnamefont {Qu\'er\'e}},
  \bibinfo {author} {\bibfnamefont {O.}~\bibnamefont {Gobert}}, \bibinfo
  {author} {\bibfnamefont {M.}~\bibnamefont {Perdrix}}, \bibinfo {author}
  {\bibfnamefont {P.}~\bibnamefont {Martin}}, \bibinfo {author} {\bibfnamefont
  {P.}~\bibnamefont {Audebert}}, \bibinfo {author} {\bibfnamefont {J.~C.}\
  \bibnamefont {Gauthier}}, \bibinfo {author} {\bibfnamefont {J.-P.}\
  \bibnamefont {Geindre}},\ and\ \bibinfo {author} {\bibfnamefont
  {T.}~\bibnamefont {Wittmann}},\ }\bibfield  {title} {\bibinfo {title}
  {Complete characterization of a plasma mirror for the production of
  high-contrast ultraintense laser pulses},\ }\href
  {https://doi.org/10.1103/PhysRevE.69.026402} {\bibfield  {journal} {\bibinfo
  {journal} {Phys. Rev. E}\ }\textbf {\bibinfo {volume} {69}},\ \bibinfo
  {pages} {026402} (\bibinfo {year} {2004})}\BibitemShut {NoStop}%
\bibitem [{\citenamefont {Amoruso}\ \emph {et~al.}(2005)\citenamefont
  {Amoruso}, \citenamefont {Ausanio}, \citenamefont {Bruzzese}, \citenamefont
  {Vitiello},\ and\ \citenamefont {Wang}}]{Amoruso}%
  \BibitemOpen
  \bibfield  {author} {\bibinfo {author} {\bibfnamefont {S.}~\bibnamefont
  {Amoruso}}, \bibinfo {author} {\bibfnamefont {G.}~\bibnamefont {Ausanio}},
  \bibinfo {author} {\bibfnamefont {R.}~\bibnamefont {Bruzzese}}, \bibinfo
  {author} {\bibfnamefont {M.}~\bibnamefont {Vitiello}},\ and\ \bibinfo
  {author} {\bibfnamefont {X.}~\bibnamefont {Wang}},\ }\bibfield  {title}
  {\bibinfo {title} {Femtosecond laser pulse irradiation of solid targets as a
  general route to nanoparticle formation in a vacuum},\ }\href
  {https://doi.org/10.1103/PhysRevB.71.033406} {\bibfield  {journal} {\bibinfo
  {journal} {Phys. Rev. B}\ }\textbf {\bibinfo {volume} {71}},\ \bibinfo
  {pages} {033406} (\bibinfo {year} {2005})}\BibitemShut {NoStop}%
\bibitem [{\citenamefont {Gattass}\ and\ \citenamefont
  {Mazur}(2008)}]{Gattass08}%
  \BibitemOpen
  \bibfield  {author} {\bibinfo {author} {\bibfnamefont {R.~R.}\ \bibnamefont
  {Gattass}}\ and\ \bibinfo {author} {\bibfnamefont {E.}~\bibnamefont
  {Mazur}},\ }\bibfield  {title} {\bibinfo {title} {Femtosecond laser
  micromachining in transparent materials},\ }\href
  {https://doi.org/10.1038/nphoton.2008.47} {\bibfield  {journal} {\bibinfo
  {journal} {Nature Photonics}\ }\textbf {\bibinfo {volume} {2}},\ \bibinfo
  {pages} {219 EP } (\bibinfo {year} {2008})}\BibitemShut {NoStop}%
\bibitem [{\citenamefont {Gamaly}(2011)}]{Gamaly11}%
  \BibitemOpen
  \bibfield  {author} {\bibinfo {author} {\bibfnamefont {E.}~\bibnamefont
  {Gamaly}},\ }\bibfield  {title} {\bibinfo {title} {The physics of ultra-short
  laser interaction with solids at non-relativistic intensities},\ }\href
  {https://doi.org/https://doi.org/10.1016/j.physrep.2011.07.002} {\bibfield
  {journal} {\bibinfo  {journal} {Physics Reports}\ }\textbf {\bibinfo {volume}
  {508}},\ \bibinfo {pages} {91 } (\bibinfo {year} {2011})}\BibitemShut
  {NoStop}%
\bibitem [{\citenamefont {Chimier}\ \emph {et~al.}(2011)\citenamefont
  {Chimier}, \citenamefont {Ut\'eza}, \citenamefont {Sanner}, \citenamefont
  {Sentis}, \citenamefont {Itina}, \citenamefont {Lassonde}, \citenamefont
  {L\'egar\'e}, \citenamefont {Vidal},\ and\ \citenamefont
  {Kieffer}}]{Chimier}%
  \BibitemOpen
  \bibfield  {author} {\bibinfo {author} {\bibfnamefont {B.}~\bibnamefont
  {Chimier}}, \bibinfo {author} {\bibfnamefont {O.}~\bibnamefont {Ut\'eza}},
  \bibinfo {author} {\bibfnamefont {N.}~\bibnamefont {Sanner}}, \bibinfo
  {author} {\bibfnamefont {M.}~\bibnamefont {Sentis}}, \bibinfo {author}
  {\bibfnamefont {T.}~\bibnamefont {Itina}}, \bibinfo {author} {\bibfnamefont
  {P.}~\bibnamefont {Lassonde}}, \bibinfo {author} {\bibfnamefont
  {F.}~\bibnamefont {L\'egar\'e}}, \bibinfo {author} {\bibfnamefont
  {F.}~\bibnamefont {Vidal}},\ and\ \bibinfo {author} {\bibfnamefont {J.~C.}\
  \bibnamefont {Kieffer}},\ }\bibfield  {title} {\bibinfo {title} {Damage and
  ablation thresholds of fused-silica in femtosecond regime},\ }\href
  {https://doi.org/10.1103/PhysRevB.84.094104} {\bibfield  {journal} {\bibinfo
  {journal} {Phys. Rev. B}\ }\textbf {\bibinfo {volume} {84}},\ \bibinfo
  {pages} {094104} (\bibinfo {year} {2011})}\BibitemShut {NoStop}%
\bibitem [{\citenamefont {Sundaram}\ and\ \citenamefont
  {Mazur}(2002)}]{Sundaram02}%
  \BibitemOpen
  \bibfield  {author} {\bibinfo {author} {\bibfnamefont {S.~K.}\ \bibnamefont
  {Sundaram}}\ and\ \bibinfo {author} {\bibfnamefont {E.}~\bibnamefont
  {Mazur}},\ }\bibfield  {title} {\bibinfo {title} {Inducing and probing
  non-thermal transitions in semiconductors using femtosecond laser pulses},\
  }\href {https://doi.org/10.1038/nmat767} {\bibfield  {journal} {\bibinfo
  {journal} {Nature Materials}\ }\textbf {\bibinfo {volume} {1}},\ \bibinfo
  {pages} {217} (\bibinfo {year} {2002})}\BibitemShut {NoStop}%
\bibitem [{\citenamefont {Tsubouchi}\ and\ \citenamefont
  {Kumada}(2012)}]{Tsubouchi12}%
  \BibitemOpen
  \bibfield  {author} {\bibinfo {author} {\bibfnamefont {M.}~\bibnamefont
  {Tsubouchi}}\ and\ \bibinfo {author} {\bibfnamefont {T.}~\bibnamefont
  {Kumada}},\ }\bibfield  {title} {\bibinfo {title} {Development of
  high-efficiency etalons with an optical shutter for terahertz laser pulses},\
  }\href {https://doi.org/10.1364/OE.20.028500} {\bibfield  {journal} {\bibinfo
   {journal} {Opt. Express}\ }\textbf {\bibinfo {volume} {20}},\ \bibinfo
  {pages} {28500} (\bibinfo {year} {2012})}\BibitemShut {NoStop}%
\bibitem [{\citenamefont {Ghimire}\ \emph {et~al.}(2011)\citenamefont
  {Ghimire}, \citenamefont {DiChiara}, \citenamefont {Sistrunk}, \citenamefont
  {Agostini}, \citenamefont {DiMauro},\ and\ \citenamefont {Reis}}]{Ghimire11}%
  \BibitemOpen
  \bibfield  {author} {\bibinfo {author} {\bibfnamefont {S.}~\bibnamefont
  {Ghimire}}, \bibinfo {author} {\bibfnamefont {A.~D.}\ \bibnamefont
  {DiChiara}}, \bibinfo {author} {\bibfnamefont {E.}~\bibnamefont {Sistrunk}},
  \bibinfo {author} {\bibfnamefont {P.}~\bibnamefont {Agostini}}, \bibinfo
  {author} {\bibfnamefont {L.~F.}\ \bibnamefont {DiMauro}},\ and\ \bibinfo
  {author} {\bibfnamefont {D.~A.}\ \bibnamefont {Reis}},\ }\bibfield  {title}
  {\bibinfo {title} {Observation of high-order harmonic generation in a bulk
  crystal},\ }\href {https://doi.org/10.1038/nphys1847} {\bibfield  {journal}
  {\bibinfo  {journal} {Nature Physics}\ }\textbf {\bibinfo {volume} {7}},\
  \bibinfo {pages} {138} (\bibinfo {year} {2011})}\BibitemShut {NoStop}%
\bibitem [{\citenamefont {Austin}\ \emph {et~al.}(2015)\citenamefont {Austin},
  \citenamefont {Kafka}, \citenamefont {Trendafilov}, \citenamefont {Shvets},
  \citenamefont {Li}, \citenamefont {Yi}, \citenamefont {Szafruga},
  \citenamefont {Wang}, \citenamefont {Lai}, \citenamefont {Blaga},
  \citenamefont {DiMauro},\ and\ \citenamefont {Chowdhury}}]{Austin15}%
  \BibitemOpen
  \bibfield  {author} {\bibinfo {author} {\bibfnamefont {D.~R.}\ \bibnamefont
  {Austin}}, \bibinfo {author} {\bibfnamefont {K.~R.~P.}\ \bibnamefont
  {Kafka}}, \bibinfo {author} {\bibfnamefont {S.}~\bibnamefont {Trendafilov}},
  \bibinfo {author} {\bibfnamefont {G.}~\bibnamefont {Shvets}}, \bibinfo
  {author} {\bibfnamefont {H.}~\bibnamefont {Li}}, \bibinfo {author}
  {\bibfnamefont {A.~Y.}\ \bibnamefont {Yi}}, \bibinfo {author} {\bibfnamefont
  {U.~B.}\ \bibnamefont {Szafruga}}, \bibinfo {author} {\bibfnamefont
  {Z.}~\bibnamefont {Wang}}, \bibinfo {author} {\bibfnamefont {Y.~H.}\
  \bibnamefont {Lai}}, \bibinfo {author} {\bibfnamefont {C.~I.}\ \bibnamefont
  {Blaga}}, \bibinfo {author} {\bibfnamefont {L.~F.}\ \bibnamefont {DiMauro}},\
  and\ \bibinfo {author} {\bibfnamefont {E.~A.}\ \bibnamefont {Chowdhury}},\
  }\bibfield  {title} {\bibinfo {title} {Laser induced periodic surface
  structure formation in germanium by strong field mid ir laser solid
  interaction at oblique incidence},\ }\href
  {https://doi.org/10.1364/OE.23.019522} {\bibfield  {journal} {\bibinfo
  {journal} {Opt. Express}\ }\textbf {\bibinfo {volume} {23}},\ \bibinfo
  {pages} {19522} (\bibinfo {year} {2015})}\BibitemShut {NoStop}%
\bibitem [{\citenamefont {Keldysh}(1965)}]{Keldysh}%
  \BibitemOpen
  \bibfield  {author} {\bibinfo {author} {\bibfnamefont {L.~V.}\ \bibnamefont
  {Keldysh}},\ }\bibfield  {title} {\bibinfo {title} {Ionization in the field
  of a strong electromagnetic wave},\ }\href@noop {} {\bibfield  {journal}
  {\bibinfo  {journal} {Sov. Phys.-JETP}\ }\textbf {\bibinfo {volume} {20}},\
  \bibinfo {pages} {1307} (\bibinfo {year} {1965})}\BibitemShut {NoStop}%
\bibitem [{\citenamefont {Runge}\ and\ \citenamefont {Gross}(1984)}]{Runge84}%
  \BibitemOpen
  \bibfield  {author} {\bibinfo {author} {\bibfnamefont {E.}~\bibnamefont
  {Runge}}\ and\ \bibinfo {author} {\bibfnamefont {E.~K.~U.}\ \bibnamefont
  {Gross}},\ }\bibfield  {title} {\bibinfo {title} {Density-functional theory
  for time-dependent systems},\ }\href
  {https://doi.org/10.1103/PhysRevLett.52.997} {\bibfield  {journal} {\bibinfo
  {journal} {Phys. Rev. Lett.}\ }\textbf {\bibinfo {volume} {52}},\ \bibinfo
  {pages} {997} (\bibinfo {year} {1984})}\BibitemShut {NoStop}%
\bibitem [{\citenamefont {Tong}\ and\ \citenamefont {Chu}(2001)}]{Tong01}%
  \BibitemOpen
  \bibfield  {author} {\bibinfo {author} {\bibfnamefont {X.-M.}\ \bibnamefont
  {Tong}}\ and\ \bibinfo {author} {\bibfnamefont {S.-I.}\ \bibnamefont {Chu}},\
  }\bibfield  {title} {\bibinfo {title} {Multiphoton ionization and high-order
  harmonic generation of he, ne, and ar atoms in intense pulsed laser fields:
  Self-interaction-free time-dependent density-functional theoretical
  approach},\ }\href {https://doi.org/10.1103/PhysRevA.64.013417} {\bibfield
  {journal} {\bibinfo  {journal} {Phys. Rev. A}\ }\textbf {\bibinfo {volume}
  {64}},\ \bibinfo {pages} {013417} (\bibinfo {year} {2001})}\BibitemShut
  {NoStop}%
\bibitem [{\citenamefont {Otobe}\ \emph {et~al.}(2008)\citenamefont {Otobe},
  \citenamefont {Yamagiwa}, \citenamefont {Iwata}, \citenamefont {Yabana},
  \citenamefont {Nakatsukasa},\ and\ \citenamefont {Bertsch}}]{Otobe08}%
  \BibitemOpen
  \bibfield  {author} {\bibinfo {author} {\bibfnamefont {T.}~\bibnamefont
  {Otobe}}, \bibinfo {author} {\bibfnamefont {M.}~\bibnamefont {Yamagiwa}},
  \bibinfo {author} {\bibfnamefont {J.-I.}\ \bibnamefont {Iwata}}, \bibinfo
  {author} {\bibfnamefont {K.}~\bibnamefont {Yabana}}, \bibinfo {author}
  {\bibfnamefont {T.}~\bibnamefont {Nakatsukasa}},\ and\ \bibinfo {author}
  {\bibfnamefont {G.~F.}\ \bibnamefont {Bertsch}},\ }\bibfield  {title}
  {\bibinfo {title} {First-principles electron dynamics simulation for optical
  breakdown of dielectrics under an intense laser field},\ }\href
  {https://doi.org/10.1103/PhysRevB.77.165104} {\bibfield  {journal} {\bibinfo
  {journal} {Phys. Rev. B}\ }\textbf {\bibinfo {volume} {77}},\ \bibinfo
  {pages} {165104} (\bibinfo {year} {2008})}\BibitemShut {NoStop}%
\bibitem [{\citenamefont {Yabana}\ \emph {et~al.}(2012)\citenamefont {Yabana},
  \citenamefont {Sugiyama}, \citenamefont {Shinohara}, \citenamefont {Otobe},\
  and\ \citenamefont {Bertsch}}]{yabana12}%
  \BibitemOpen
  \bibfield  {author} {\bibinfo {author} {\bibfnamefont {K.}~\bibnamefont
  {Yabana}}, \bibinfo {author} {\bibfnamefont {T.}~\bibnamefont {Sugiyama}},
  \bibinfo {author} {\bibfnamefont {Y.}~\bibnamefont {Shinohara}}, \bibinfo
  {author} {\bibfnamefont {T.}~\bibnamefont {Otobe}},\ and\ \bibinfo {author}
  {\bibfnamefont {G.~F.}\ \bibnamefont {Bertsch}},\ }\bibfield  {title}
  {\bibinfo {title} {Time-dependent density functional theory for strong
  electromagnetic fields in crystalline solids},\ }\href
  {https://doi.org/10.1103/PhysRevB.85.045134} {\bibfield  {journal} {\bibinfo
  {journal} {Phys. Rev. B}\ }\textbf {\bibinfo {volume} {85}},\ \bibinfo
  {pages} {045134} (\bibinfo {year} {2012})}\BibitemShut {NoStop}%
\bibitem [{\citenamefont {Sato}\ \emph {et~al.}(2015)\citenamefont {Sato},
  \citenamefont {Yabana}, \citenamefont {Shinohara}, \citenamefont {Otobe},
  \citenamefont {Lee},\ and\ \citenamefont {Bertsch}}]{Sato15}%
  \BibitemOpen
  \bibfield  {author} {\bibinfo {author} {\bibfnamefont {S.~A.}\ \bibnamefont
  {Sato}}, \bibinfo {author} {\bibfnamefont {K.}~\bibnamefont {Yabana}},
  \bibinfo {author} {\bibfnamefont {Y.}~\bibnamefont {Shinohara}}, \bibinfo
  {author} {\bibfnamefont {T.}~\bibnamefont {Otobe}}, \bibinfo {author}
  {\bibfnamefont {K.-M.}\ \bibnamefont {Lee}},\ and\ \bibinfo {author}
  {\bibfnamefont {G.~F.}\ \bibnamefont {Bertsch}},\ }\bibfield  {title}
  {\bibinfo {title} {Time-dependent density functional theory of high-intensity
  short-pulse laser irradiation on insulators},\ }\href
  {https://doi.org/10.1103/PhysRevB.92.205413} {\bibfield  {journal} {\bibinfo
  {journal} {Phys. Rev. B}\ }\textbf {\bibinfo {volume} {92}},\ \bibinfo
  {pages} {205413} (\bibinfo {year} {2015})}\BibitemShut {NoStop}%
\bibitem [{\citenamefont {Noda}\ \emph {et~al.}(2018)\citenamefont {Noda},
  \citenamefont {Sato}, \citenamefont {Hirokawa}, \citenamefont {Uemoto},
  \citenamefont {Takeuchi}, \citenamefont {Yamada}, \citenamefont {Yamada},
  \citenamefont {Shinohara}, \citenamefont {Yamaguchi}, \citenamefont {Iida},
  \citenamefont {Floss}, \citenamefont {Otobe}, \citenamefont {Lee},
  \citenamefont {Ishimura}, \citenamefont {Boku}, \citenamefont {Bertsch},
  \citenamefont {Nobusada},\ and\ \citenamefont {Yabana}}]{salmon}%
  \BibitemOpen
  \bibfield  {author} {\bibinfo {author} {\bibfnamefont {M.}~\bibnamefont
  {Noda}}, \bibinfo {author} {\bibfnamefont {S.~A.}\ \bibnamefont {Sato}},
  \bibinfo {author} {\bibfnamefont {Y.}~\bibnamefont {Hirokawa}}, \bibinfo
  {author} {\bibfnamefont {M.}~\bibnamefont {Uemoto}}, \bibinfo {author}
  {\bibfnamefont {T.}~\bibnamefont {Takeuchi}}, \bibinfo {author}
  {\bibfnamefont {S.}~\bibnamefont {Yamada}}, \bibinfo {author} {\bibfnamefont
  {A.}~\bibnamefont {Yamada}}, \bibinfo {author} {\bibfnamefont
  {Y.}~\bibnamefont {Shinohara}}, \bibinfo {author} {\bibfnamefont
  {M.}~\bibnamefont {Yamaguchi}}, \bibinfo {author} {\bibfnamefont
  {K.}~\bibnamefont {Iida}}, \bibinfo {author} {\bibfnamefont {I.}~\bibnamefont
  {Floss}}, \bibinfo {author} {\bibfnamefont {T.}~\bibnamefont {Otobe}},
  \bibinfo {author} {\bibfnamefont {K.-M.}\ \bibnamefont {Lee}}, \bibinfo
  {author} {\bibfnamefont {K.}~\bibnamefont {Ishimura}}, \bibinfo {author}
  {\bibfnamefont {T.}~\bibnamefont {Boku}}, \bibinfo {author} {\bibfnamefont
  {G.~F.}\ \bibnamefont {Bertsch}}, \bibinfo {author} {\bibfnamefont
  {K.}~\bibnamefont {Nobusada}},\ and\ \bibinfo {author} {\bibfnamefont
  {K.}~\bibnamefont {Yabana}},\ }\bibfield  {title} {\bibinfo {title} {Salmon:
  Scalable ab-initio light--matter simulator for optics and nanoscience},\
  }\bibfield  {journal} {\bibinfo  {journal} {Computer Physics Communications}\
  }\href {https://doi.org/https://doi.org/10.1016/j.cpc.2018.09.018}
  {https://doi.org/10.1016/j.cpc.2018.09.018} (\bibinfo {year}
  {2018})\BibitemShut {NoStop}%
\bibitem [{\citenamefont {Bertsch}\ \emph {et~al.}(2000)\citenamefont
  {Bertsch}, \citenamefont {Iwata}, \citenamefont {Rubio},\ and\ \citenamefont
  {Yabana}}]{Bertsch00}%
  \BibitemOpen
  \bibfield  {author} {\bibinfo {author} {\bibfnamefont {G.~F.}\ \bibnamefont
  {Bertsch}}, \bibinfo {author} {\bibfnamefont {J.-I.}\ \bibnamefont {Iwata}},
  \bibinfo {author} {\bibfnamefont {A.}~\bibnamefont {Rubio}},\ and\ \bibinfo
  {author} {\bibfnamefont {K.}~\bibnamefont {Yabana}},\ }\bibfield  {title}
  {\bibinfo {title} {Real-space, real-time method for the dielectric
  function},\ }\href {https://doi.org/10.1103/PhysRevB.62.7998} {\bibfield
  {journal} {\bibinfo  {journal} {Phys. Rev. B}\ }\textbf {\bibinfo {volume}
  {62}},\ \bibinfo {pages} {7998} (\bibinfo {year} {2000})}\BibitemShut
  {NoStop}%
\bibitem [{\citenamefont {Becke}\ and\ \citenamefont
  {Johnson}(2006)}]{Becke2006}%
  \BibitemOpen
  \bibfield  {author} {\bibinfo {author} {\bibfnamefont {A.~D.}\ \bibnamefont
  {Becke}}\ and\ \bibinfo {author} {\bibfnamefont {E.~R.}\ \bibnamefont
  {Johnson}},\ }\bibfield  {title} {\bibinfo {title} {A simple effective
  potential for exchange},\ }\href {https://doi.org/10.1063/1.2213970}
  {\bibfield  {journal} {\bibinfo  {journal} {The Journal of Chemical Physics}\
  }\textbf {\bibinfo {volume} {124}},\ \bibinfo {pages} {221101} (\bibinfo
  {year} {2006})},\ \Eprint
  {https://arxiv.org/abs/https://doi.org/10.1063/1.2213970}
  {https://doi.org/10.1063/1.2213970} \BibitemShut {NoStop}%
\bibitem [{\citenamefont {Tran}\ and\ \citenamefont {Blaha}(2009)}]{mBJ2}%
  \BibitemOpen
  \bibfield  {author} {\bibinfo {author} {\bibfnamefont {F.}~\bibnamefont
  {Tran}}\ and\ \bibinfo {author} {\bibfnamefont {P.}~\bibnamefont {Blaha}},\
  }\bibfield  {title} {\bibinfo {title} {Accurate band gaps of semiconductors
  and insulators with a semilocal exchange-correlation potential},\ }\href
  {https://doi.org/10.1103/PhysRevLett.102.226401} {\bibfield  {journal}
  {\bibinfo  {journal} {Phys. Rev. Lett.}\ }\textbf {\bibinfo {volume} {102}},\
  \bibinfo {pages} {226401} (\bibinfo {year} {2009})}\BibitemShut {NoStop}%
\bibitem [{\citenamefont {Perdew}\ and\ \citenamefont {Zunger}(1981)}]{LDA}%
  \BibitemOpen
  \bibfield  {author} {\bibinfo {author} {\bibfnamefont {J.~P.}\ \bibnamefont
  {Perdew}}\ and\ \bibinfo {author} {\bibfnamefont {A.}~\bibnamefont
  {Zunger}},\ }\bibfield  {title} {\bibinfo {title} {Self-interaction
  correction to density-functional approximations for many-electron systems},\
  }\href {https://doi.org/10.1103/PhysRevB.23.5048} {\bibfield  {journal}
  {\bibinfo  {journal} {Phys. Rev. B}\ }\textbf {\bibinfo {volume} {23}},\
  \bibinfo {pages} {5048} (\bibinfo {year} {1981})}\BibitemShut {NoStop}%
\bibitem [{\citenamefont {Troullier}\ and\ \citenamefont
  {Martins}(1991)}]{TM91}%
  \BibitemOpen
  \bibfield  {author} {\bibinfo {author} {\bibfnamefont {N.}~\bibnamefont
  {Troullier}}\ and\ \bibinfo {author} {\bibfnamefont {J.~L.}\ \bibnamefont
  {Martins}},\ }\bibfield  {title} {\bibinfo {title} {Efficient
  pseudopotentials for plane-wave calculations},\ }\href
  {https://doi.org/10.1103/PhysRevB.43.1993} {\bibfield  {journal} {\bibinfo
  {journal} {Phys. Rev. B}\ }\textbf {\bibinfo {volume} {43}},\ \bibinfo
  {pages} {1993} (\bibinfo {year} {1991})}\BibitemShut {NoStop}%
\bibitem [{\citenamefont {Kleinman}\ and\ \citenamefont
  {Bylander}(1982)}]{Kleinman82}%
  \BibitemOpen
  \bibfield  {author} {\bibinfo {author} {\bibfnamefont {L.}~\bibnamefont
  {Kleinman}}\ and\ \bibinfo {author} {\bibfnamefont {D.~M.}\ \bibnamefont
  {Bylander}},\ }\bibfield  {title} {\bibinfo {title} {Efficacious form for
  model pseudopotentials},\ }\href
  {https://doi.org/10.1103/PhysRevLett.48.1425} {\bibfield  {journal} {\bibinfo
   {journal} {Phys. Rev. Lett.}\ }\textbf {\bibinfo {volume} {48}},\ \bibinfo
  {pages} {1425} (\bibinfo {year} {1982})}\BibitemShut {NoStop}%
\end{thebibliography}%
\end{document}